# Quantum Wave Resistance of Schrodinger functions

Timir Datta

Physics & Astronomy Department and the Nanocenter,

University of South Carolina

&

Raphael Tsu

Electrical Engineering Department,

University of North Carolina at Charlotte

### Overview

The wave particle duality of the quantum mechanical description of matter is well known; here we report a hitherto unknown wave property of the Schrodinger function. Wave resistance is fundamental to all classical waves. Is there also impedance to a quantum-wave, $Z(k,\omega)$ or a quantum wave - impedance (QWI)? QWI will be an analogue of Maxwell's free space impedance ($Z_{EM} = \sqrt{(\mu_o/\varepsilon_o)}$) where, $\varepsilon_0$ and $\mu_0$ are the electromagnetic constants.

We show, the impedance, $Z^{lmn}$, for a Schrodinger, free-particle, $\psi(r,t)_{lmn}$ is non-zero, purely real (resistive) and can be expressed in terms of the fine structure constant ($\alpha$) as, $Z_{lmn} = (\pi\sqrt{(\mu_o/\varepsilon_o)}/\alpha)\, \Xi_{lmn}$, per spin. $Z_{lmn}$ is the determinant of the flow and partitioning of charge and energy transported by the system and numerically its scale is ($h/2q^2 \sim 12.9 k\Omega$), so the corresponding wave-conductance $G_0$ (77.5 µmho, per spin) is double the unit of Landaur conductance. As functions of the quantum numbers ($l, m, n$) the geometric factor, $\Xi_{lmn}$ shows, peaks, valleys and plateaus; also, as in quantum hall-effect, both integer and fractional values viz., 5/3, 5/2, 13/5, etc., in 2-dimensions, and 3/2, 9/5, 11/5, etc., in 3-dimensions. Furthermore, these steps occur without the presence of any magnetic field. In microwaves and optics, $Z(k,\omega)$ is essential in impedance matching for effective power transfer, the impedance defined here is of no exception. Quantum wave-impedance offers a unified approach to a wide range of fundamental and technological situations.

## Introduction

Energy of a wave and its flow are intriguing topics even for classical fields (Feynman, Leighton and Sands 1965). For example, in the complete absence of scattering, there is intrinsic wave-resistance or impedance (Z). Z is a measure of the transfer of energy concomitant to wave propagation and is fundamental to all waves (Lamb 1932, Morse and Feshbach 1953). Ideal waves in strings, on the surfaces of a liquid or those inside solids, all have impedance. Maxwell's electromagnetic waves traveling in perfect vacuum show impedance of 376.7Ω (Stratton 1941), the intrinsic free space impedance, $Z_{EM} = \sqrt{\mu_o/\varepsilon_o}$ where, $\varepsilon_0$ and $\mu_0$ are respectively the permittivity and permeability constants. In contrast particles are different; for instance in the particle description, resistance to an electrical current arises from friction due to scattering of the carriers; with zero scattering, currents propagate without impediment.

In quantum mechanics any object is described by its wave function, $\psi(r, t)$, (Dirac 1958, Landau and Lifshitz 1977, and Kroemer 1994). The wave function, $\psi(r,t)$ is generally a complex quantity, which represents the probability amplitude of the system and as such not physically measurable. However, the works by Aharonov, Bhom (Aharanov & Bohm, 1959), Berry (Berry, 1984 and 1990) and others clearly show the wave geometry and observational reality of $\psi(r,t)$. Processes such as tunnelling, (Tsu and Esaki 1973) are practical technologically important consequences of quantum waves. A travelling $\psi(r,t)$ satisfies the customary equation of wave propagation, viz., $\zeta = f(x \pm vt)$. As the electron travels it carries energy and charge along with it.

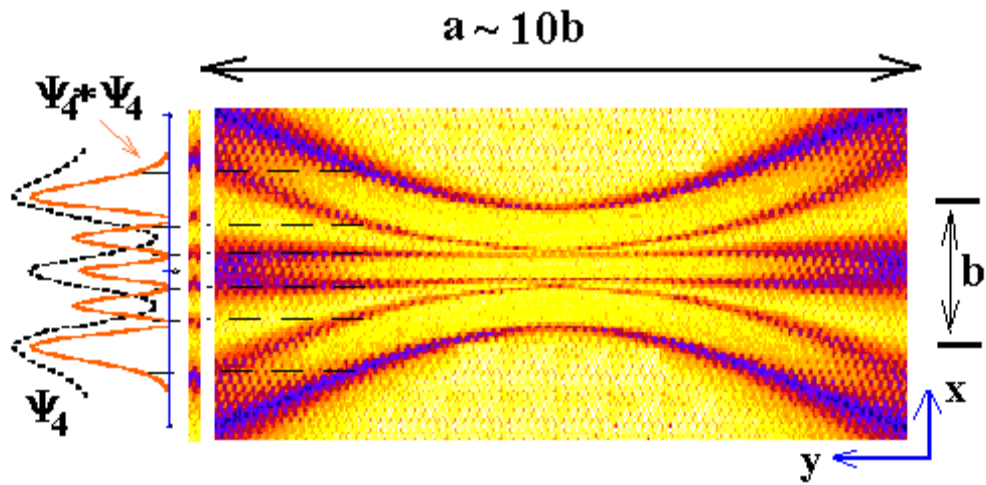

**Figure 1**- the coherent, time independent, wake patterns of a travelling electron wave, in the vicinity of a narrow gate. A comparison with the ideal fringe pattern for the $4^{th}$ excited state expected for a harmonic potential, $U(x) \sim (1/2)kx^2$, is shown on the left. Clearly, U over the entire picture depends on both x and y, also it extends beyond the electrodes and opens up away from them, giving rise to the hour glass shape of the wakes observed above.

Since the advent of scanning probe microscopy the position probability density, $\psi(r,t)^*\psi(r,t)$, particularly those associated with electrons have been imaged extensively; the "quantum corral" is a notable example of electron standing waves in copper. Figure 1 is a rendition of an electron wave flowing through a narrow gate region. The potential profile, $U(x)$ in the gate region, is approximately harmonic in the (x) direction perpendicular to the propagation (y) direction. A fit of the fringe pattern with the ideal behaviour expected of the $4^{th}$ harmonic oscillator state, is indicated on the left. The steady state nodal pattern of $\psi(r,t)^*\psi(r,t)$ along the direction of propagation are clearly visible. However, the travelling wave fronts perpendicular to motion cannot be observed because these are rapidly oscillating and are averaged out during the time take to make the image. The coherence of the wake persists over a considerable region.

**Theory**

In this communication, we ask the questions, is there impedance to a propagating Schrodinger-wave, $\psi(\mathbf{k},\omega)$ and how to calculate it? The idea of pure quantum wave impedance, (QWI), was first introduced by Tsu (Tsu 2002 and 2003). When current, I, is measured in amperes and the electric potential (V) or energy per unit charge is measured in volts then resistance or impedance $Z(k,\omega)$ is in ohms ($\Omega$), and its inverse conductance $G(k,\omega)$ is measured in mhos ($\Omega^{-1}$). In a quantum mechanical system, of charge, q, with Hamiltonian, H, and carrying a net (irrotational) current, we define QWI by the following vector relationship (Datta & Tsu, 2003):

$$q\int (J.ds) \bullet Z(k,\omega) = \frac{1}{q}\int (\psi * H\psi dV) \qquad (1a)$$

Or as a scalar by the quotient[‡]:

$$Z(k,\omega) = \frac{\frac{1}{q}\int (\psi * H\psi dV)}{q\int (J.ds)} \qquad (1b)$$

Like wise we reason the following relation holds for the corresponding QWC that is the quantum wave conductance, $G(k,\omega)$:

$$G\left(\iiint \psi * H\psi dv\right) = q^2 \left(\iint J.ds\right) \qquad (2)$$

The V in above the equations is the volume of the system. The q times J, integrated over the transporting surface is equal to the net current I. Where, J, the probability current density is defined by the equation,

$$\frac{\partial(\psi *\psi)}{\partial t} + \nabla \bullet J = 0 \qquad (3)$$

The electron carries charge and energy as it moves forward along the propagating wave vector **k**. An electron that carries a lot of energy but a small amount of current would have large impedance (small conductance) and vice versa. These ideas hold as long as the phase coherence of the wave is retained all over the volume of the wave. In the case of scattering or phase coherence loss, the definitions may be generalized such as by taking the appropriate ensemble average. Considerations of dimensional analysis require that the natural unit or scale factor for all quantum resistance (impedance) be the quantity $h/q^2$ (25.8kΩ) where h is Planck's constant.

The Hamiltonian influences the resistance (or conductance) in several of ways (Datta 2003), the obvious being the relationship explicit in equations 1 & 2. The other is via current density J through the time derivative in equation 3, because the Hamiltonian operator controls time evolution. In the classical limit, our definitions reduce to their corresponding versions of the traditional wave impedance. Also, because these equations (1-2) state the relationship between the rate of flow rate and energy, they are like quantum mechanical analogues of Ohm's law ♣. However, unlike the classical Ohm's law, wave impedance (conductance) does not arise from a dissipative, frictional flow.

Next let us go over some simple but illustrative calculations of QWI and QWC. As our first example we will consider the case of a spin-less, "free particle" of effective mass $m_o$. The relevant $H^§$, $\psi(k,\omega)$ and J are given by the well know expressions (Dirac 1958, Landau & Lifshitz 1977, and Kroemer 1994):

$$H = \frac{-(\frac{h}{2\pi})^2}{2m_o}\nabla^2 \quad (4a)$$

$$\psi(k,\omega) = \left(\frac{1}{\sqrt{V}}\right)\exp[i(k_x x + k_y y + k_z z - \omega t)] \quad (4b)$$

and

$$J = \frac{-i(\frac{h}{2\pi})}{2m_o}[\psi^*(\nabla\psi) - \nabla(\psi^*)\psi] \quad (4c)$$

Identifying the state by the quantum numbers, *l, m, n,* and one obtains,

$$Z_{lmn} = \frac{1}{2}(\frac{h}{q^2})\Xi_{lmn} \quad (5)$$

Notice, for this case of a single effective mass $m_o$, $Z_{lmn}$ is independent of $m_o$. As with quantum hall effect (QHE), equation 6a may be recast to incorporate the Sommerfeld-Dirac fine structure constant, $\alpha$ ($2\pi q^2/hc \sim 1/137$), viz.,

$$Z_{lmn} = \frac{1}{2}(\frac{2\pi\sqrt{\varepsilon_o\mu_o}}{\alpha})\Xi_{lmn} \quad (6a)$$

We can also cast QWI in term of the electromagnetic free-space impedance, (Stratton 1941) $Z_{EM}$ as follows,

$$Z_{lmn} = \frac{1}{2}(\frac{c^2\mu_o}{2\alpha})Z_{EM}\Xi_{lmn} = Z_0\Xi_{lmn} \qquad (6b)$$

$Z_{lmn}$ as given by equations 6a-6b is independent of the effective mass of the particle. The pre-factor $Z_0$ is equal to $h/2q^2$. The half in $Z_0$ arises from the electron group velocity. The values of the impedance function $\Xi_{lmn}$, i.e., $Z_{lmn}/Z_0$ are typically non-zero. Hence even for a perfectly collision free, coherent motion with no irreversibility (Landaur 1970 and 1989) a finite value of the impedance (resistance) ensues. This observation about finite impedance pertains only to equations. 6a-6b (the pure wave contribution) and not in general to the other forms of resistance (impedance), such as those arising from boundaries, contacts, interfaces, reflections and scatterings (Buttiker 1989, Datta 1995).

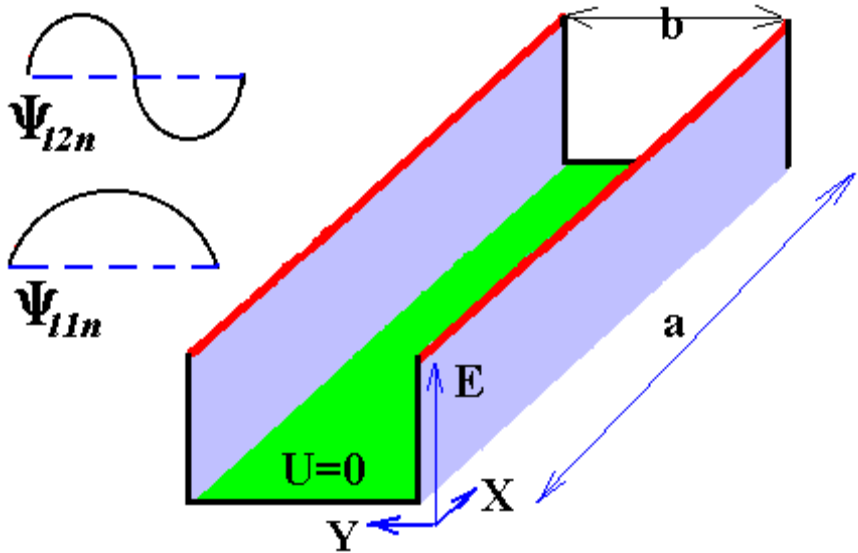

**Figure 2** – schematics of a 2-dimensional quantum well with "hard" walls. The potential U is zero in the channel, which is shaded pale green; the (light purple) walls are at infinite potentials. The y-dependence of the real parts of the two lowest *m* states (*l*,1 & *l*,2) are indicated on the upper left corner.

The functional form of $\Xi_{lmn}$, depends on geometric and quantization conditions imposed on the system. For an electron wave-guide of dimensions a, b and c, with edges aligned along the x, y, and z, axes respectively and with one dimensional wave propagation along the x-axis, i.e., standing waves in the transverse directions ($J_y$ & $J_z$ =0) two hard boundary conditions are required. A simple two-dimensional well structure may be seen in figure 2. In this case the potential is zero in the channel and abruptly, becomes infinite on both the walls. Ballistic propagation requires that $\zeta \geq a$, where, $\zeta$ is the electron coherence length and $\lambda$ the de Broglie wavelength.

For analytical simplicity we will consider two sets of eigensolutions. In a given frame of reference, we first consider integer wavelengths in size "a" along the propagation direction, then:

$$k_x a = l(2\pi)$$
$$k_y b = m\pi$$
$$k_z c = n\pi$$

(7a)

Where, *l, m, n,* are positive integers. The real parts of the standing wave pattern along the y-axis, of two lowest "*m*" wave functions *l,1(n)* & *l,2 (n)* are indicated on the upper left corner of figure 2. The relative probability distribution, $\Psi^*\Psi$ (at time, $t=t_0$) of the *l*=4 and *m*=3 state for $\lambda_x < \lambda_y$ are shown in figure 3. Where the distance between the next nearest spots on the probability plot, along each of the coordinate axis, is the corresponding wavelength. The wave vector **k** relates to the inverse space, i.e., $k_x = 2\pi/\lambda_x$ and $k_y = 2\pi/\lambda_y$. Also shown is the real part of the wave function $\Psi(x,y,t_0)$. At all times, energy E, and current **I** propagate parallel to **k**. $Z_{lmn}$, is a measure of the quantum mechanical energy carried by an electron per unit electric current. With time, both the top and bottom figures coherently translate along the direction of **k**. Under suitable boundary conditions the loci of the probability distribution of such a travelling wave may produce wakes patterns that are fixed in space, similar to those shown in figure 1.

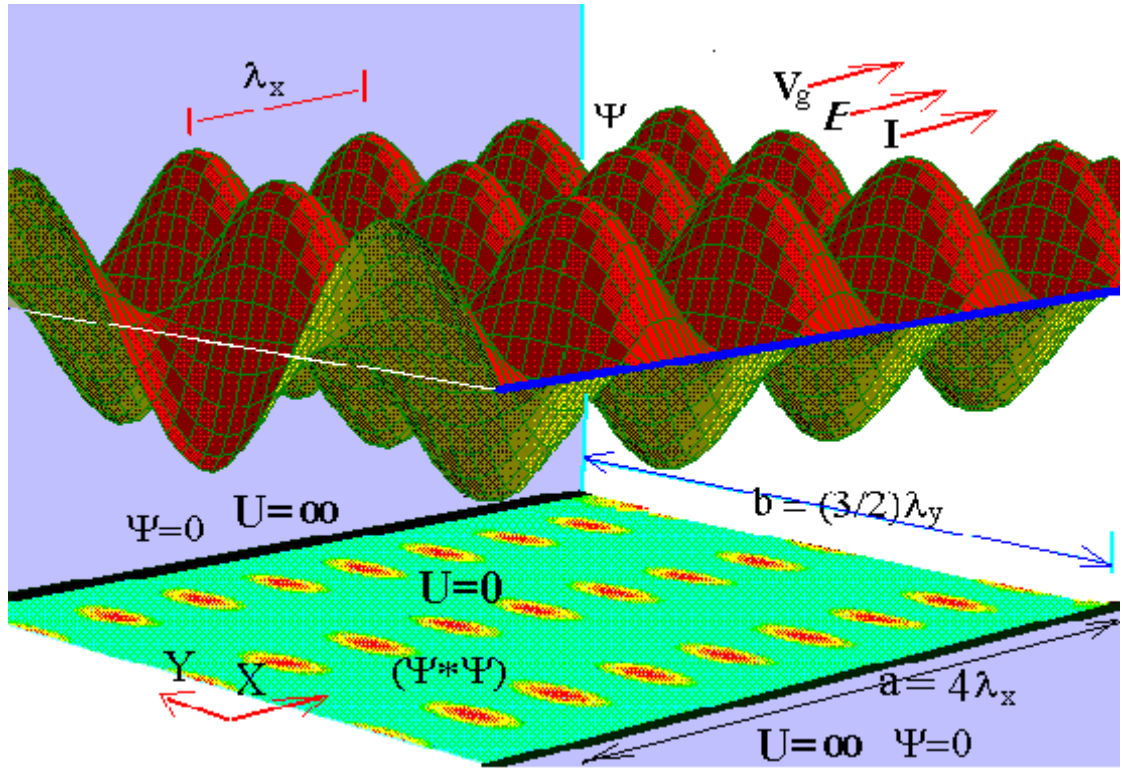

**Figure 3**- The instantaneous x and y dependence of the real part of a 2-dimensional, free-electron wave function $\psi(x,y,t_0)$ for the state $l = 4$ and $m = 3$ is shown by the top surface. On the bottom is a projection of the corresponding relative electron probability distribution. The direction of the energy flow (E), group velocity ($V_g$) and that of current (I) are indicated in the top right corner.

Conditions (7a) lead to an expression for $Z_{lmn}$, which is akin to the generalized, microwave guide formula, namely:

$$\Xi_{lmn}(a,b,c;l,m,n) = \left(\frac{a^2}{l}\right)\left[\frac{l^2}{a^2} + \frac{1}{4}\left(\frac{m^2}{b^2} + \frac{n^2}{c^2}\right)\right] \tag{7b}$$

Second is the Ramsauer (anti-reflective) condition, or when integer numbers of half-wave lengths equal the size "a" then:

$$k_x a = l(\pi)$$
$$k_y b = m\pi$$
$$k_z c = n\pi \tag{7c}$$

Conditions (7c) lead to the following expression for $Z_{lmn}$:

$$\Xi_{lmn}(a,b,c;l,m,n) = \frac{1}{2}\left(\frac{a^2}{l}\right)\left[\frac{l^2}{a^2} + \frac{m^2}{b^2} + \frac{n^2}{c^2}\right] \tag{7d}$$

Characteristic of the increased transmission when half integer wavelengths fit in length "a", the value of impedance given by equation 7d is less than that of equation 7b. In the general case of a three-dimensional travelling wave, as may be realized in the vacuum gap region between nanoscale electrodes, periodic quantization conditions prevail, in all three directions:

$$k_x a = l(2\pi)$$
$$k_y b = m(2\pi)$$
$$k_z c = n(2\pi) \tag{8a}$$

As in equation 7a the quantum numbers in equation 8a are positive integers. The impedance function, corresponding to equation 8a is:

$$\Xi_{lmn}(a,b,c;l,m,n) = \left(\frac{\left[\frac{l^2}{a^2}+\frac{m^2}{b^2}+\frac{n^2}{c^2}\right]}{\left[\frac{l}{a^2}+\frac{m}{b^2}+\frac{n}{c^2}\right]}\right) \quad (8b)$$

Our second example will be to calculate the QWC of a free electron in a cylindrical shell of length L and radii $R_a$ and $R_b$ (figure 4).

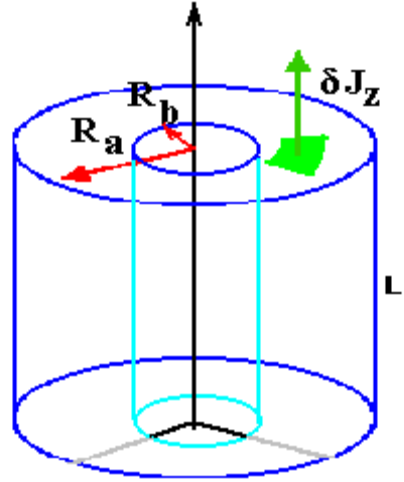

**Figure 4-** A cylinder of length L and radii $R_a$ and $R_b$. The centre of the bottom may be taken as the origin of a Cartesian coordinate system so that J is parallel to the z axis and the elemental surface area ds (green) lies in the z = L plane.

In this case of a perfectly conducting, cylinder the following equations hold,

$$H = [\frac{p^2}{2m_0}] = [\frac{-\hbar^2}{2m_0}\nabla_{cyl}^2] = [\frac{-\hbar^2}{2m_0}\{\partial_z^2 + \partial_\rho^2\} + \frac{L_z^2}{2m_0\rho^2}] \quad (9a)$$

$$\psi = A_{lmn} J_m(K_{mn}\rho)\exp[i(m\phi + k_l z - \omega t)] \quad (9b)$$

Because, the cylindrical shell is bounded by $R_a$ and $R_b$ the zeros of the Bessel functions will have to be at both the inner and the outer surfaces imposing standing wave conditions on the cylinder surfaces, namely,

$$J_m(K_{mn}R_a) = J_m(K_{mn}R_b) = 0 \quad (9c)$$

Again one can consider the two cases, (i) L equal to integer wavelengths and (ii) the resonant condition of L equal to half integer wavelengths. For the first case:

$$k_l L = l(2\pi) \qquad (9d)$$

In the case of a solid cylinder, $R_a = R_b = R$ and we have:

$$\iint I \cdot ds = \iint I_z d\phi \rho d\rho = \left(\frac{\hbar}{2m_0 R}\right) 2\pi A_{lmn}^2 [\int J_{mn}(K_{mn}\rho)^2 \rho d\rho](k_z R)$$

$$= \left(\frac{\hbar}{m_0 RL}\right)(l2\pi) = l\left(\frac{h}{m_0 RL}\right)$$

and

$$\iiint \Psi^* H \Psi d\rho d\phi dz = \left(\frac{\hbar^2(K_{mn}^2 + k_l^2)}{2m_0}\right) = \left(\frac{\hbar^2 k_l^2(1 + K_{mn}^2/k_l^2)}{2m_0}\right)$$

$$= \hbar^2(l2\pi)^2 \left(\frac{(1 + K_{mn}^2/k_l^2)}{2m_0 R^2}\right) = \left(\frac{(hl)^2}{2m_0}\right)\frac{(1 + K_{mn}^2/k_l^2)}{R^2} = \left(\frac{(hl)^2}{2m_0}\right)[\frac{\{1 + (K_{mn}R)^2/(k_l R)^2\}}{R^2}]$$

$$= \left(\frac{h^2}{2m_0 R^2}\right) l^2 \gamma_{lmn}(R)$$

$$(9e)$$

Where

$$\gamma_{lmn}(R) = [\{1 + (K_{mn}R)^2\}/(k_l R)^2]$$

$$G_{lmn} = \left(\frac{q^2 \iint J d\rho d\phi}{\iiint \Psi^* H \Psi d\rho d\phi dz}\right) = \frac{q^2\left(\frac{lh}{m_0 RL}\right)}{\left(\frac{h^2(l)^2}{2m_0 R^2}\right)\gamma_{lmn}(R)} = \frac{2q^2}{hl}[\gamma_{lmn}(R)\frac{L}{R}]^{-1}$$

Or,

$$G_{lmn}(R,L) = G_0 g_{lmn}(R,L) = \frac{1}{Z_0}[l\frac{L}{R}\gamma_{lmn}(R)]^{-1} \quad (9f)$$

Equation 9f, defines the geometric factor $g_{lmn}$, $G_0$ is $1/Z_0$ and as before $Z_o$ is equal to $h/(2q^2)$ or 12.9kΩ. Not surprisingly G is proportional to the ratio of the radius and length and once again independent of the effective mass. G for different values of the parameters are best calculated numerically and will be reported elsewhere.

**Discussion of Results**

To get a physical insight into $Z_{lmn}$ let us discuss results from the simple case of a cube, i.e., a= b =c, so that from equation 8b one obtains:

$$\Xi_{lmn} = \left[\frac{l^2 + m^2 + n^2}{l + m + n}\right] \quad (10a)$$

Equation 10a follows a non-monotonic dependence on its arguments. As long as the electron is in the $l^{th}$ state i.e., $l=m=n$, $\Xi_{lll}$ is the same for all dimensions and it follows:

$$Z_l = l[\frac{1}{2}(\frac{h}{q^2})] = lZ_0 \quad (10b)$$

and the corresponding conductance is given by

$$G_l = \frac{1}{Z_l} = (Z_0 l)^{-1} \quad (10c)$$

Where $Z_1 = Z_0$ and the conductance is $G_1$, $Z_0^{-1} = 2q^2/h$. Numerically the unit of wave–conductance, $G_0$ is 77.6 μmho per spin, this is exactly two times that of "Landaur conductance" (Webb et al. 1985). Conductors with two units of conductance steps are observed and have been reported (Javery et al., 2003). Also, $l$ is a quantum number not the "channel number".

For a given wave, equation 9b shows the $l$ dependence of Z, for a given "device" with a fixed geometry. As $l$ increases, energy goes up faster than current and as a net result impedance increases linearly with $l$. With energy degeneracy, the current at a given energy is increased by the contributions from each of the degenerate channels (states). For example with two spin states, the current will be two times larger and the corresponding conductance will be doubled. Like wise in a solid, G has to be multiplied by the number of valley degeneracy along the direction of transport.

Some of the values of $\Xi_{lmn}$, (calculated from equation 8b) for one-, two- and three-dimensions are listed in Table-1. Here repeated quantum number combinations are entered only once. A series of numbers including integers and fractions are possible. For example: 5/3, 5/2, 13/5, 17/5, etc., in 2-d and 3/2, 9/5, 11/5, 7/3, 17/7 and others in 3-d are noted.

Integer and fractional steps such as these are novel quantum features and are strikingly reminiscent of integer QHE (v. Klitzing et al 1980) and fractional QHE (Tusi et al. 1982). Remarkably, these steps arise purely from the geometry of the problem and do not require the strong correlation or filling of landau levels.

**Table-1**

Quantum number dependence of the impedance function $\Xi_{lmn}$ in 1,2 & 3 Euclidean dimensions, the values up to $l$=4 are shown.

| $\ell$ | $m$ | $n$ | 1-d | 2-d | 3-d |
|---|---|---|---|---|---|
| 1 | 1 | 1 | 1 | 1 | 1 |
| 2 | 1 | 1 | - | 5/3 | 3/2 |
| 2 | 2 | 1 | - | - | 9/5 |
| 2 | 2 | 2 | 2 | 2 | 2 |
| 3 | 1 | 1 | - | 5/2 | 11/5 |
| 3 | 2 | 1 | - | - | 7/3 |
| 3 | 2 | 2 | - | 13/5 | 17/7 |
| 3 | 3 | 1 | - | - | 19/7 |
| 3 | 3 | 2 | - | - | 11/4 |
| 3 | 3 | 3 | 3 | 3 | 3 |
| 4 | 1 | 1 | - | 17/5 | 3 |
| 4 | 2 | 1 | - | - | 3 |
| 4 | 2 | 2 | - | 10/3 | 3 |
| 4 | 3 | 1 | - | - | 13/4 |
| 4 | 3 | 2 | - | - | 29/9 |
| 4 | 3 | 3 | - | 25/7 | 17/5 |
| 4 | 4 | 1 | - | - | 11/3 |
| 4 | 4 | 2 | - | - | 18/5 |
| 4 | 4 | 3 | - | - | 41/11 |
| 4 | 4 | 4 | 4 | 4 | 4 |

.

Low quantum number regions for several constant values of *l* are plotted in figure 5. Peaks, valleys as well as plateaus appear in the $\Xi_{lmn}$ -landscape. The plateaus represent 'degeneracies', in the quotient, $Z_{lmn}/Z_o$. For example, in three dimensions, several combinations of adjacent quantum numbers viz., 3,3,3; 4,1,1; 4,2,1 & 4,2,2; give rise to the same value of $\Xi_{lmn}$ = 3.

As noted above, in a real system the ideal-ballistic regime is restricted to system sizes less than the electron coherence length ($\zeta$). Finite values of $\zeta$ will set limits to the range of quantum numbers and applicability of the free-QWI calculation.

It may be useful to compare QWI (QWC) with other forms of impedance. For example, the well-known Landaur formula (Landaur 1989) calculates G as transmission coefficients ($T_{ij}$) weighted by the quantum scale factor of conductance ($q^2/h$) namely,

$$G = \left(\frac{q^2}{h}\right)\sum T_{ij} \qquad (11)$$

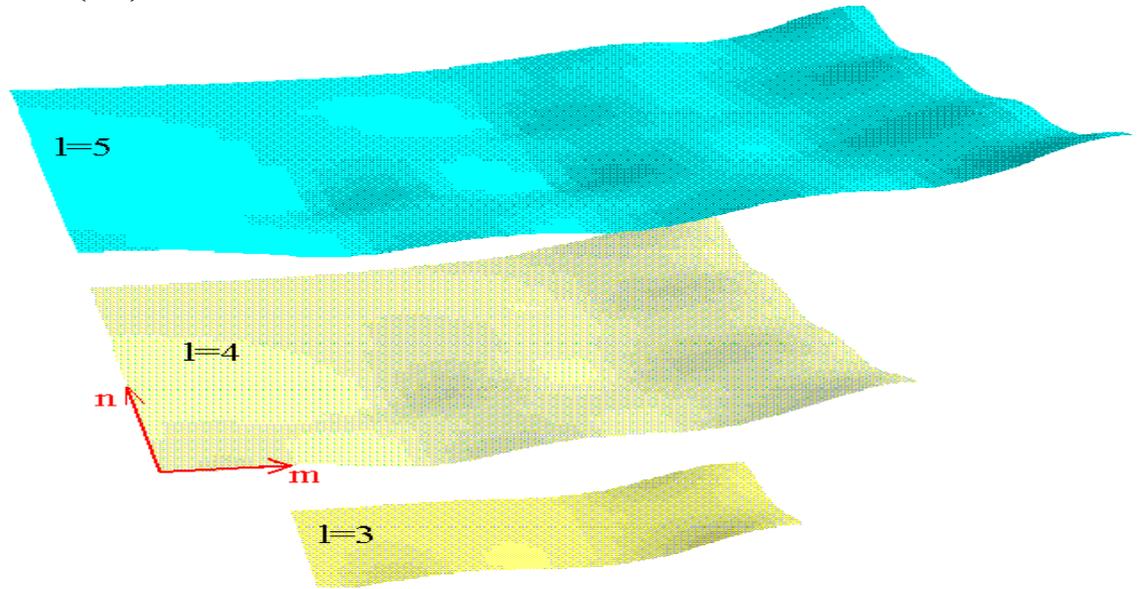

**Figure 5-** The $\Xi_{lmn}$ surfaces equation 8b for three constant values of *l* (3,4,5) are shown. Notice the up and down topography of $\Xi_{lmn}$ in the quantum number landscape.

Where the sum extends over all the propagating channels including spin. The transmission coefficients are defined as the ratio of the transmitted and incident current densities. Equation 11 is a measure of how well the current flows irrespective of the waves (explicit) energy content.

In contrast QWI measures the correlated flow of energy with charge. This distinction may be made clear by considering the wave in a narrow channel such as in the infinite well structure (figure 2) associated with equation 8b. Of the two wave functions *l,1 (n)* & *l,2 (n)* shown, the state with m=2 has higher energy. Both of these states have the same $k_x$, the same transmission coefficients and the same current but different energies and different $Z_{lmn}$!

Conventional or classical wave-resistance, $Z(k,\omega)$, is predicated by the flow of energy. Due to its association with oscillatory energy, $Z(k,\omega)$ is indispensable for the determination of power transfer and impedance matching. Not surprisingly Maxwell's free-space impedance is a critical engineering parameter in telecommunication and in general $Z(k,\omega)$ has played a pivotal role in the development of microwave technology. The QWI defined here is likely to be no exception in future technologies.

## Summary and Conclusions

The parameters quantum wave impedance (QWI) and conductance (QWC) were introduced to determine if there is wave resistance to a perfectly collision less, free-particle wave function. QWI is a relationship between charge and energy flow rate that is valid in any coherent system as long as there is energy and a flow of a quantum attribute, viz., charge, spin, and others energy and unlike the Ohm's law equation, wave impedance (conductance) is not because of dissipation or friction.

We have shown that for a free-particle Schrodinger, wave function $\psi(r,t)_{lmn}$, the quantum wave impedance $Z_{lmn}$ is non-zero, quantized and purely real or resistive. It is related to the fine structure constant, $\alpha$ and $\varepsilon_o$ and $\mu_o$, the electromagnetic constants of free-space and $Z_{lmn}$ is equal to $(\pi\sqrt{(\varepsilon_o\mu_o)}/\alpha)\Xi_{lmn}$, per spin. Where, $\Xi_{lmn}$ is a geometric factor. Numerically the quantum or the scale of $Z_{lmn}$ is equal to $h/2q^2$ ~12.9k$\Omega$, so the corresponding "wave-conductance" $G_0$ ~77.5 µmho (per spin).

Even more intriguing, as functions of the quantum numbers *(l, m & n)*, $\Xi$ shows, peaks, valleys and plateaus. Also similar to quantum hall-effect (QHE) both integer and fractional values viz., 5/3, 5/2, 13/5, etc., in 2-dimensions, and 3/2, 9/5, 11/5, etc., in 3-dimensions are allowed. Interestingly, these integer and fractional values of QWI take place for free-particle states without any magnetic field. Although, we have explicitly considered Schrodinger functions, our general conclusions should hold for any coherent quantum wave.

Due to its association with the flow of oscillatory energy classical wave impedance is an indispensable engineering parameter. It has also played a pivotal role in the development of power and communication technology. The quantum impedance, $Z_{lmn}$ defined here is likely to be no exception. In addition it may provide a new standard of high-precision metrology. Also, quantum wave impedance offers a unified approach to a wide range of fundamental and technological questions.

## Acknowledgements

The first author would like to acknowledge partial supports from the Naval Research Laboratory and the USC Nano-center. The second author would like to acknowledge the US Army Research Office for the support throughout the years of his research.

**Footnotes:**

‡ By the correspondence principle equations 1a & 1b reduce to the generalized definitions of the customary impedance valid for mechanical and electromagnetic waves. However, we may also define a dynamical, $z = \frac{1}{q}(\frac{\partial H}{\partial I})$ .

♣ We can obtain the canonical kinetic theory result for conductivity σ =$ne^2t/m$ from equation 1b, by considering a cube with sides of length l if each particle propagates at the average speed of (v/2) and $1/2mv^2$ as the energy transported (Datta 2003).

§ This is written in a gauge where the constant potential energy of the free particle is zero.